# *AssessITS:* Integrating procedural guidelines and practical evaluation metrics for organizational IT and Cybersecurity risk assessment

Mir Mehedi Rahman, School of Business & Technology, Emporia State University, mrahman2@g.emporia.edu
Naresh Kshetri, Department of Cybersecurity, Rochester Institute of Technology, naresh.kshetri@rit.edu
Sayed Abu Sayeed, School of Computing, University of South Alabama, ss2352@jagmail.southalabama.edu
Md Masud Rana, School of Business & Technology, Emporia State University, mrana@emporia.edu

**Abstract** - In today's digitally driven landscape, robust Information Technology (IT) risk assessment practices are essential for safeguarding systems, digital communication, and data. This paper introduces 'AssessITS', an actionable method designed to provide organizations with comprehensive guidelines for conducting IT and cybersecurity risk assessments. Drawing extensively from NIST 800-30 Rev 1, COBIT 5, and ISO 31000, 'AssessITS' bridges the gap between high-level theoretical standards and practical implementation challenges. The paper outlines a step-by-step methodology that organizations can simply adopt to systematically identify, analyze, and mitigate IT risks. By simplifying complex principles into actionable procedures, this framework equips practitioners with the tools needed to perform risk assessments independently, without too much reliance on external vendors. The guidelines are developed to be straightforward, integrating practical evaluation metrics that allow for the precise quantification of asset values, threat levels, vulnerabilities, and impacts on confidentiality, integrity, and availability. This approach ensures that the risk assessment process is not only comprehensive but also accessible, enabling decision-makers to implement effective risk mitigation strategies customized to their unique operational contexts. 'AssessITS' aims to enable organizations to enhance their IT security strength through practical, actionable guidance based on internationally recognized standards.

## I. INTRODUCTION

In a time where information technology (IT) is the foundation of nearly all business operations, effectively managing IT-related risks is crucial for maintaining organizational stability and security. Effective management of IT risk is of utmost importance. However, research indicates that conventional frameworks continue to have challenges in appropriately addressing the intricate nature of IT hazards [1]. IT risks involve uncertainties arising from insufficient information about threats, advancing technologies, and internal organizational vulnerabilities, among other issues. This paper presents a holistic approach to assessing risks called 'AssessITS'. It combines step-by-step directions with realistic evaluation metrics to improve the efficiency of IT risk management techniques. This method not only seeks to detect and reduce risks but also prepares organizations to adjust to various operating situations, thereby strengthening their ability to withstand future IT and cybersecurity interruptions.

---

The study innovates by combining theoretical frameworks such as NIST 800-30 Rev.1, COBIT 5, and ISO 31000 into a practical, customizable methodology. While each of these frameworks addresses specific approaches for risk assessment, *'AssessITS'* streamlines them into a unified approach. This integration simplifies complex IT risk assessments by providing clear, actionable steps and metrics that make it easier to adopt across various industries and organizational sizes. Compared to existing standards that often require substantial interpretation, *'AssessITS'* offers a more straightforward path from theory to practical implementation. However, these standards were selected for this study due to their global recognition in IT risk management and governance. NIST 800-30 focuses on risk assessment processes, COBIT 5 on IT governance and management, and ISO 31000 provides comprehensive guidelines for risk management. Together, they create a balanced and robust framework suited for assessing IT and cybersecurity risks.

Organizations have widely embraced standardized IT management frameworks, which have firmly established certain risk management processes. These frameworks promote approaches that are widely recognized as 'best practices' because of their proven efficacy in many industries [2]. Nevertheless, these techniques frequently face criticism due to their limited capacity to adjust to the distinct environmental and operational obstacles that are particular to each firm [2]. This assessment is especially applicable in dynamic sectors where technical threats and company strategies undergo fast evolution. Although these frameworks are commonly used, they often lack a strong mechanism for continuously adapting to new hazards. Instead, they rely on fixed models that do not accurately represent the dynamic nature of the business environment.

The incorporation of IT governance frameworks, such as COBIT, is essential in harmonizing IT and business plans to optimize the generation of value from IT expenditures. These frameworks offer organized models that not only improve the administration of IT resources but also guarantee that IT services are in line with the strategic requirements of the business. Organizations may ensure efficient management of IT assets and alignment with business objectives by adopting a holistic approach that incorporates governance and management [3]. This approach supports overall corporate governance and strategic alignment. Alignment is crucial in the current digital environment, where IT plays a major role in attaining business agility and operational success.

The expansion of digital infrastructure necessitates a heightened focus on the interaction between IT management frameworks and dynamic risk settings. Although frameworks such as COBIT provide standardized methodologies that are crucial for aligning IT operations with business objectives, they still lack the ability to adapt and respond in real time to new risks [4]. Moreover, the dynamic and ever-changing nature of cybersecurity risks requires more than fixed models; it requires a framework that not only predicts possible weaknesses but also incorporates ongoing learning and adjustment procedures. Research suggests that there is a significant variation in the effectiveness of current risk management methods, especially when it comes to dealing with the complexities of organizations and technology [5]. 'AssessITS' tackles these difficulties by offering a versatile, data-driven approach that connects theoretical concepts with practical execution, thereby strengthening organizational resilience against IT and cybersecurity threats.

The goal of the study is to break down high-level guidelines into specific, manageable tasks. For organizations of all sizes and industries, it provides evaluation metrics and procedural guidelines that simplify risk assessments, offering clarity in implementation, ensuring even smaller organizations can apply this matrix effectively. This strategy offers the potential to not only simplify risk assessment operations but also enhance the strategic integration of IT risk assessment with overall company objectives.

## II. RELATED WORK

In risk management, regardless of the domain, whether healthcare, financial, IT, or other sectors, the fundamental approach to assessing and mitigating risks tends to follow similar methodologies. Angel Rodriguez-Calero et al. [5] explored risk factors related to "difficult peripheral venous access" in healthcare, using a systematic process to identify and evaluate risks based on factors like obesity and gender. This approach, like many in risk management, involves analyzing potential threats, assessing their likelihood, and determining the impact, which mirrors the structure of a typical risk matrix. No matter the field, risk assessments mostly rely on categorizing and prioritizing risks based on severity and probability, using a consistent framework to guide decision-making and mitigation strategies.

European Union Agency for Cybersecurity (ENISA) [6] is an agency dedicated to achieving a high level of cybersecurity across Europe. The purpose is to provide a coherent overview of published standards that describe methodologies and address aspects of risk management to confirm with these standards. It can be used by organizations as a library of risk management standards to develop cybersecurity certification or implement risk management within their organizations. Network Information Security Directive 2016/1148 (NIS) provides several references on risk management as the first cybersecurity legislation. European institutions have released several publications including regulations and directives - regarding security of ICT, crucial EU legal requirements are presented in Cybersecurity Act Regulation - Regulation (EU) 2019/881 (CSA regulation).

M. S. Saleh & A. Alfantookh [7] presented a comprehensive Information Security Risk Management (ISRM) framework that enables the efficient establishment of a safe target environment as use of e-transactions in enterprise is becoming necessary. The framework uses the various assessment tools along with six sigma cycles - define, measure, analyze, improve, and control (DMAIC) for the ISRM process. Use of "support tools" like collection and survey tools, mathematical modeling tools, software packages, other related rolls etc. for various phase processes is considered as is the case with other ISRM methods. While assessment criteria is considered open to various standards, the framework uses the comprehensive view - strategy, technology, organization, people, and environment (STOPE) for the ISRM scope.

T. Aven [8] performed a review on advances in risk assessment and risk management. Author reflected on where future development of the risk field is needed, encouraged and also looked for trends in perspectives and approaches. Main conclusions are (i) Scientific foundation is shaky on some issues as both theory and practice rely on decision-makers perspectives, (ii) Attempts at integrated research are being conducted with broader perspectives on risk management, assessment, conceptualization in recent years, and (iii) For meeting the challenges faced in risk field related to societal problems there are signs of revitalization of interests in foundational issues in risk assessment and management. As methods exist and are developing, to provide important contributions in decision-making practice, risk assessment and risk management are required in the scientific field.

O. Gadyatskaya et. al. [9] proposed the TRICK security risk assessment methodology to bridge the gap between two worlds by introducing optimal countermeasure selection problems on attack-defense trees. Risk treatment relies on catalogs of countermeasures and analysts are expected to estimate the residual risks according to ISO/IEC 27001. The gap between practical risk assessment and academic research because practical risk assessment methods do not include state-of-the-art scientific results. Practical risk assessment methodologies often have a wider scope than specific academic developments leading to an interfacing problem which is another possible factor. Authors provided a special algorithm that has been tested and implemented in cloud security context and a high-level description of the proposed extension of TRICK.

**III. RISK ASSESSMENT CONCEPTS**

Risk assessment is an integral part of a comprehensive risk management process and is necessary for efficient information security management in businesses. This process entails a methodical analysis of possible risks to an organization's operations, assets, and personnel, along with the discovery of weaknesses from both internal and external sources. The process involves assessing the negative consequences that may occur if these threats take advantage of known weaknesses and determining the probability of such events. This allows the organization to make well-informed decisions about how to respond to risks in accordance with its overall risk management strategy [10]. The approach highlights the importance of risk assessment in preserving the organization's security and incorporates it into the overall risk management framework, encouraging a proactive approach to addressing potential security threats. The comprehensive risk management framework commences with risk framing, which establishes the foundation by describing the environmental backdrop and delineating the methodology for subsequent risk-related choices [11]. The risk assessment phase involves the identification and evaluation of threats and vulnerabilities. This step is crucial as it provides information for the decision-making process by calculating the possible impact and likelihood of adverse events [10]. Afterward, the risk response component is initiated, in which the business formulates and implements solutions specifically designed to reduce the impact of identified risks while adhering to the predetermined risk tolerance [11]. Implementing a proactive response strategy is crucial for ensuring organizational resilience against potential threats. Finally, the risk monitoring phase guarantees the efficiency of risk management endeavors across time, adjusting to modifications in the operational environment and aligning with the overarching organizational objectives and compliance prerequisites [11]. Ongoing surveillance is crucial for continuously improving risk tactics and guaranteeing the organization's long-term security well-being.

Effective risk management in the dynamic field of cybersecurity relies on sophisticated and flexible risk assessment approaches. The highlighted multicriteria decision framework shows a sophisticated way to assess cybersecurity risks by comprehensively evaluating multiple risk components. This approach systematically measures threats, vulnerabilities, and potential consequences, combining them using a decision-analytic approach to successfully prioritize risk management methods. By evaluating risks across various dimensions, including physical, informational, and sociocognitive aspects, it represents the complex interconnections that are typical of contemporary cyber systems. The approach's primary advantage resides in its capacity to not only detect and examine stationary risks but also to adjust to the ever-changing nature of cybersecurity concerns. It assists decision-makers by offering a clear and organized procedure that corresponds to corporate objectives and technological criteria, enabling a strategic reaction to cyber threats. The incorporation of multicriteria analysis into cybersecurity risk assessments is a substantial improvement for overall risk management. This integration enables a more comprehensive comprehension of threats and promotes the creation of stronger cybersecurity systems [12].

The dynamic modeling of cyber risk assessment demonstrates how strategic evaluations of investments in cybersecurity can assist decision-making, especially for small enterprises. This study utilizes the CENSOR model to systematically examine the economic efficiency of different cybersecurity initiatives, offering a nuanced perspective on the trade-off between cost and risk reduction. This strategy is particularly beneficial for small enterprises, as they often have limited resources and must make careful decisions about where to use their funds. The framework utilizes optimization models such as Set Covering and Knapsack issues to help firms develop cybersecurity solutions that are both cost-efficient and highly effective in reducing risk exposure. The CENSOR framework assists small businesses in making optimal decisions to improve their cybersecurity posture without overspending. It achieves this by quantitatively evaluating the relationship between investment intensity and risk reduction. The framework aligns with industry standards and academic research on risk management best practices [13]. The CENSOR model demonstrates how the model offers a practical approach for small enterprises to make informed, cost-effective decisions by utilizing systematic, data-driven risk assessments to optimize cybersecurity investments and effectively reduce risks. It directly relates to the core focus on optimizing cybersecurity risk management.

Another case study highlights the crucial importance of top management in leading and implementing successful cybersecurity risk assessments, particularly following major cybersecurity incidents. The study demonstrates that when breach costs are high, the Top Management Team (TMT) becomes more focused on cybersecurity, which has a substantial impact on the organization's strategic response through Information Security Risk Assessments (ISRA). The findings emphasize the crucial role of the TMT in both managing immediate reactions after a breach and guiding the firm toward a stronger cybersecurity position. The TMT's mediation guarantees that the cybersecurity measures are not solely responsive but are integrated into the organization's strategic framework, in line with long-term business goals and regulatory requirements. An essential strategy for firms seeking to improve their ability to withstand future cybersecurity attacks is to adopt a proactive and dynamic risk management approach [14].

A holistic strategy combining qualitative and quantitative approaches is essential when assessing risk assessment ideas for varied industries such as aviation systems and medium-sized businesses. The significance of complex, multi-faceted risk assessment methods in detecting, analyzing, and reducing cyber threats is highlighted in several industries [15][16]. It is critical that these risk assessment frameworks are designed to be flexible, allowing for the incorporation of emerging security measures and technologies. This flexibility is essential for dealing with new dangers like cryptojacking and fileless ransomware, which might change and go unnoticed [17]. Existing risk assessment frameworks urgently need to be expanded to accommodate the changes blockchain technology is bringing to IT infrastructures across various sectors, including healthcare, banking, defense models, and academia [18] [19]. To guarantee effective security and integration across all platforms, these frameworks should incorporate both conventional IT setups and the distinctive features of blockchain technology. The adoption of successful techniques in holistically managing IT risks is fostered by this holistic approach, which increases the design of robust cybersecurity solutions adapted to unique industrial concerns.

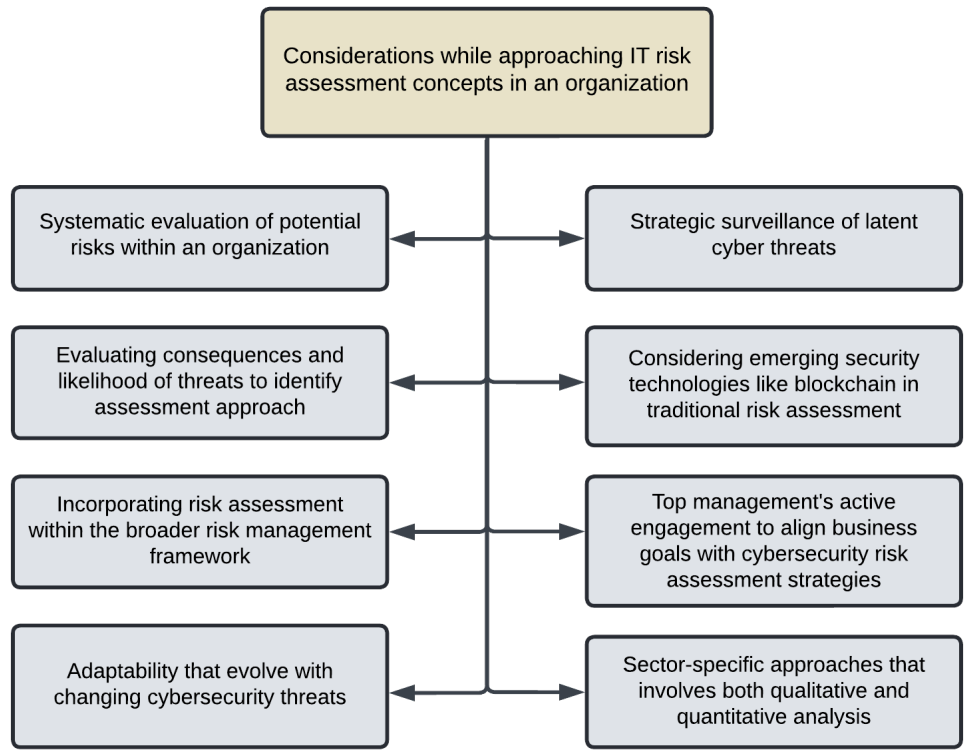

Figure 1: Considerations while approaching risk assessment concepts in an organization [10] - [18]

## IV. ESTABLISHING THE FOUNDATION FOR IT RISK ASSESSMENTS

In the field of IT and cybersecurity in organizations, practitioners typically follow a structured and comprehensive approach when building the foundation of risk assessments. This approach involves four key steps: preparing for the assessment, conducting the assessment, communicating the outcomes, and maintaining the assessment results over time [10]. The assessment process starts with an overview that establishes the fundamental framework for effectively engaging with the risk environment. Preparation activities are crucial to ensure a comprehensive state of readiness. These activities are followed by the systematic execution of the assessment, which aims to capture and analyze relevant risk data. Consequently, this section will explore the methods for effectively communicating and sharing the outcomes of the risk assessment to ensure that all stakeholders are well-informed and actively involved. Finally, the section emphasizes the significance of consistently maintaining the outcomes to accurately represent the changing risk environment. This is achieved through the use of comprehensive risk tables and assessment scales that assist in standardizing the evaluation process [20]. The researchers describe each phase through specific tasks, with additional guidance provided for organizations to perform risk assessments effectively, illustrating the basic steps and highlighting the tasks involved [20].

### 4.1 Prepare and Align Assessment Plan

It is important to prepare for an IT risk assessment by clearly defining the scope, boundaries, and context of the risk management processes. This involves ensuring a thorough understanding of all assets, systems, and data that need to be protected. This step is crucial in order to fully understand all critical assets, systems, and data [20]. By utilizing frameworks such as COBIT 2019, researchers can enhance the organization of this phase by customizing IT governance practices to align with the specific needs of the organization [21]. In order to achieve successful implementation, it is important for organizations to involve stakeholders to ensure that the risk assessment is in line with the business goals. It is also crucial to clearly define the scope of the assessment, including all assets, data, and systems. Additionally, reviewing existing security controls to identify any weaknesses is essential. Establishing clear risk tolerance levels to guide mitigation efforts is another important step. Lastly, utilizing frameworks such as NIST and COBIT can help in organizing the risk assessment process. This integrates security and privacy systematically, aligning IT risk management with strategic business objectives [20][21].

## 4.2 Risk Assessment Execution

Implementing the risk assessment entails a thorough examination to document and prioritize information security risks according to their severity and potential impact on the organization [10]. The research process begins by identifying the assets, processes, or services that will be assessed. The intrinsic value of these assets is then evaluated, along with the threats that are associated with them [20]. The severity of each identified threat is analyzed, and the impacts on the confidentiality, integrity, and availability (CIA) of information systems are assessed for vulnerabilities [10]. The assessment process involves the computation of threat values, which integrate levels of threats and vulnerabilities, as well as the determination of the probability of these threats demonstrating [10]. A risk impact rating is determined by multiplying the value of the asset, the value of the threat, and the likelihood, which helps in determining the level of risk criticality [10]. When considering the management of election risks, it is important to acknowledge that risk assessments can become more complex. This complexity arises from the need to address uncertainties and the dynamics of stakeholder reactions. It is worth noting that outcomes can vary significantly due to the introduction of new information or changes in the environment [22]. This approach ensures that risk assessments conducted by researchers are comprehensive, align with the organizational risk management framework, and are tailored to the specific definitions, guidance, and direction established during the preparatory phase, ensuring full coverage of the threat space within the constraints of available resources [20][22].

## 4.3 Communication & Information Sharing

One of the most important steps in the risk assessment process is the third step, which involves effectively communicating the results and sharing risk-related information throughout the organization. This step ensures that decision-makers have the necessary information to make informed risk decisions [10]. The security personnel need to undertake two specific tasks: firstly, communicating the outcomes of the risk assessment clearly and comprehensively, and secondly, disseminating the information gathered during the assessment to support other essential risk management activities [10]. It is important to provide a thorough and comprehensive description of this communication. This includes presenting the findings and methodologies in a way that is clear and easily comprehensible for all stakeholders involved [20]. Within the realm of Internet of Medical Things (IoMT) devices, it is crucial for the risk assessor to recognize the intricate nature of the risk landscape, which is primarily influenced by the diverse range of technologies utilized. Consequently, it becomes imperative to tailor communication strategies in order to effectively cater to the varying levels of understanding among different stakeholders. It is important to ensure that the information shared is not only comprehensible but also actionable, facilitating effective risk management decisions across diverse departments and teams [23]. Through the strategic integration of structured communication and information-sharing tasks, organizations have the potential to greatly improve the transparency and effectiveness of their risk management processes. Adopting a proactive and well-informed strategy is crucial for researchers to effectively manage and reduce potential risks in a constantly changing risk environment [10][20][23].

## 4.4 Maintaining the Risk Assessment

Maintaining a risk assessment should be a continuous process for every organization, as it allows them to effectively adjust their risk management strategies to the ever-changing landscape of threats, technological advancements, and evolving research practices. It is important to conduct periodic reviews and updates to ensure that the assessments remain up-to-date and accurately reflect the most recent security and privacy challenges [20]. The organizational practice of conducting risk assessments every six months demonstrates a systematic approach to proactively identify and address potential vulnerabilities. Regular updates to the assessment and correction of identified risks using new data need to be practiced routinely. Additionally, organizations need to verify the effectiveness of implemented risk mitigation strategies. This process assists in adapting to changes in the organization's information systems and the environments in which they function, guaranteeing continuous compliance and alignment with the organization's risk tolerance [20][10]. Through the consistent maintenance of an up-to-date risk assessment, organizations can acquire valuable insights that enable them to make informed decisions. These decisions not only improve the overall security posture of organizations but also ensure compliance with regulatory requirements. Additionally, by proactively managing security and privacy risks, organizations can enhance their long-term operational resilience and contribute to their strategic success.

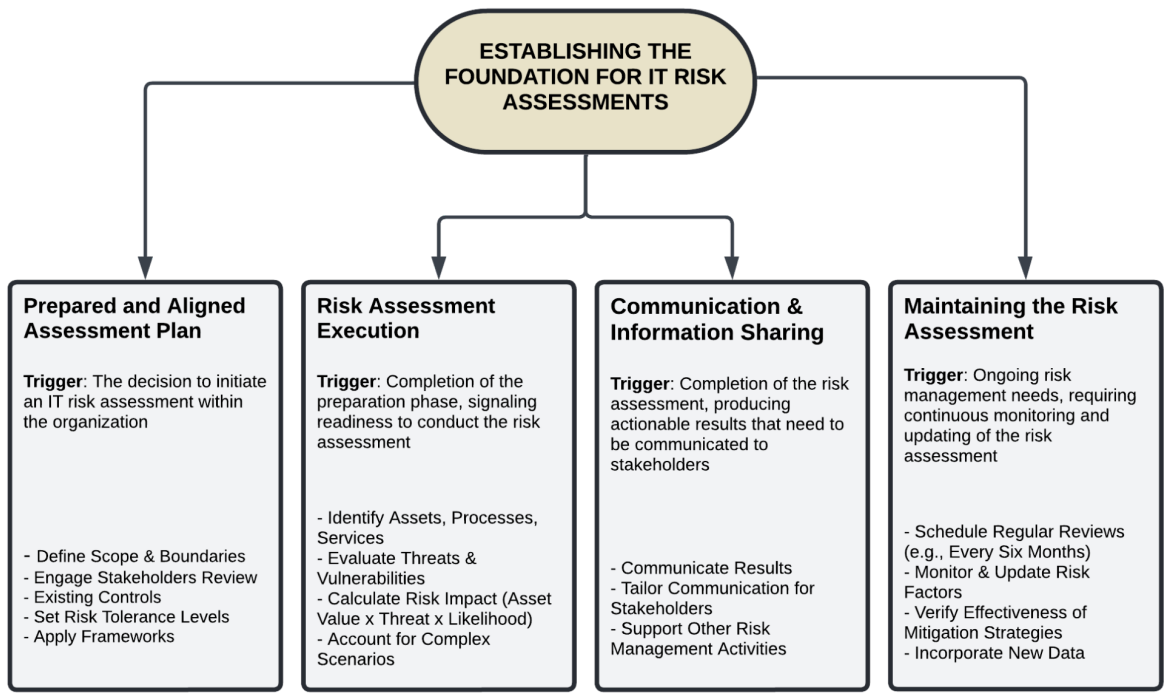

Fig 2: Establishing the foundation for IT risk assessment [10] [20] [21] [22] [23]

## V. CONDUCTING COMPREHENSIVE RISK ASSESSMENTS

With the emergence of rapid and high volume of disruptive innovation, the risk associated with it is also significantly increasing day by day. High investment in IT can never be fruitful for an organization until there is a strong IT governance system. Risk assessment is one of the strong tools of IT governance. Nowadays, risk assessment is conducted not only in IT sectors but also in other areas such as food safety, production, supply chain, physical energy systems, and so on. Because of the unique structure and complex operational process, risk managers must conduct a suitable risk assessment that brings a positive outcome for the organization. This portion of the paper will describe the clear idea about conducting risk assessment, which will include risk management and risk assessment concepts along with key risk concepts such as risk models, assessment approaches, and analysis approaches. The latter portion of this section will provide some robust ideas about the application of risk assessment.

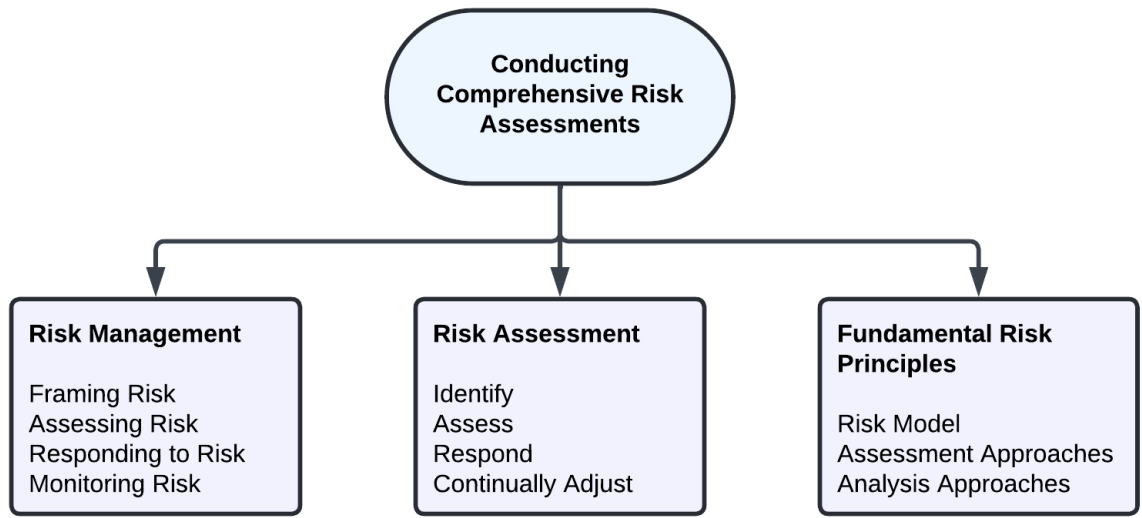

Figure 3: Conducting Comprehensive Risk Assessment [10]

## 5.1 RISK MANAGEMENT

The concepts of risk assessment, science, and risk management are always complicated, and it is getting more and more complex day by day just because of disruptive communication methods, suspicion of authorities, and increasing demands for public participation in the decision-making process [24]. According to NIST, risk management is a continuing process that manages risks in different areas of an organization, such as operations, assets, or individuals, where the risk assessment is a complementary component of risk management. There are four components of the risk management process: (1) framing risk, (2) assessing risk, (3) responding to risk, and (4) monitoring risk [10].

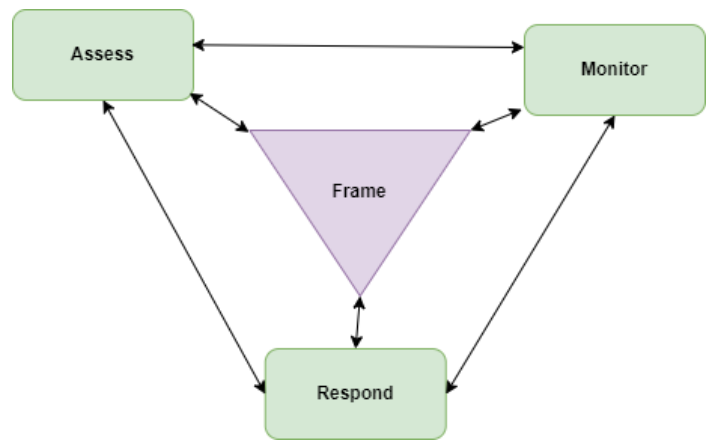

Figure 4: Risk assessment within risk management [10]

The risk framing components generate overall risk management strategies addressing how to assess, respond, and monitor risk. The risk assessment component identifies several areas, such as possible threats to the farm, internal and external vulnerabilities of the farm, potential negative impact from vulnerabilities, and the possibility of threats that will occur. The risk response component delivers continuous organizational-wide response by keeping pace with the risk frame. Lastly, the risk monitoring component determines the efficiency of the response and verifies the proper implementation of planned risk response [10].

**5.2 RISK ASSESSMENT**

According to the definition of NIST, Risk assessment is one of the significant tools of the risk management process [10]. Devos stated that risk assessment is one of the most significant scientific tools for conducting risk analysis in the food safety area [24]. According to Kandasami, Risk assessment is the fundamental segment of the risk management process that identifies risks associated with all of the organizational assets and makes risk estimation and prioritization [25]. From the definition above, it can be clearly stated that Risk assessment is an integral part of the Risk Management process, which is a step-by-step process that executes the risk analysis. The main focus of risk assessment is to identify how to conduct the assessment, how to prepare for the assessment, how to utilize the result of the assessment, and finally, how to continue the assessment over a period of time.

**5.3 Fundamental Risk Principles**

With the rapid growth of information and communication technology, uncertainty and threats are following a very significant upward trend. The probable loss associated with that uncertainty and threat is known as risk. Risk is the measurement of possible loss that can occur in the future due to unwanted circumstances. The estimation of probable loss and the probability of the occurrence of unwanted circumstances are two integral parts of risk. The term risk is part and parcel of every business organization. There is no single organization or department that is free of risk. The idea of risks varies from department to department and organization to organization. In the case of information and communication technology, monetary or reputational losses regarding data breaches, loss of data, and security breaches are some examples of risk.

To conduct a proper risk assessment, it is crucial to follow robust risk assessment methodologies that closely align with the structure of the business. Risk assessment methodologies are built on statistical or probabilistic analyses that assess the effect of adverse circumstances on the organization [27]. Risk assessment, risk model, assessment approach, and analysis approach are some essential parts of risk assessment methodologies [10].

**5.3.1 Risk Models**

Generally, a risk model is a mathematical representation of the factors that are considered in risk assessment. A risk model also tries to identify the relationship and impacts among the factors. The risk factor varies from organization to organization. Some of the vital risk factors that are common in almost every organization are vulnerability, threat, probability, impact, and predisposing condition [10]. Risk models can be very different from organization to organization. The risk model for finance is not similar to the model of accounting. Similarly, the risk models are not exactly the same in every IT department. Radanliev developed a strong model to measure the highest possible loss over a given time period, focusing on the impact of IoT on the economy [26], whereas Kandasamy developed a risk model focusing on IoT using a framework known as CORAS, which is UML based [25]. Even if both of them focus on the risk assessment of IoT, their risk models are different due to the differences in organizational structure and objective.

One of the vital components of a risk model is threat, which means any event or situation with probable adverse impact on the organization. There are a variety of threats to an organization. If we think about an IT department, server vandalism, unwanted power outages, security breaches, outages due to natural calamities, destruction, and unauthorized access are some examples of threats that may cause some adverse impact on the IT department. Likelihood, another component of the risk model, refers to the probability of occurrence of a threat. This weighted probability is identified by considering several historical factors and rigorous analysis. The impact component of the risk model identifies the level of impact on the operations or activities of an organization due to the occurrence of a threat. Another component of the risk model is vulnerability, which means the potential weakness of a system, governance, or implementation that might be impacted by the possible sources of threats [10].

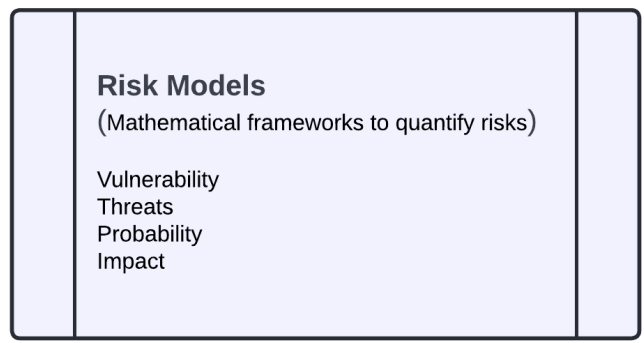

Fig 5: Risk Models [10]

### 5.3.2 Assessment Approaches

With the growing complexity of disruptive innovation, it becomes extremely difficult for organizations to follow a single approach to make the risk assessment. There is no perfect risk assessment approach in the modern world. Organizations need to select one approach from the set of alternatives that best aligns with their objective and organizational culture. NIST suggests three robust approaches for risk assessment, which are known as 1) Quantitative Approach, 2) Qualitative Approach, and 3) Semi-quantitative approach [10].

Approaches suggested by NIST have some advantages and disadvantages. If the question arises in mind about which one is perfect for a particular organization, the only answer to this question will be, "Select one approach that best aligns with your objective and organizational environment." It is important for an organization to pick the right assessment approach to get an effective outcome. Selecting the wrong approach can be devastating and can lead the organization toward failure.

The quantitative approach harnesses the principles, rules, and Techniques based on numbers to make the risk assessment. This approach is widely accepted and used so far, especially in cost-benefit analysis. In this approach, the factors or properties of the factors are converted into numbers. Afterwards, those numbers are processed and calculated following the set of principles and guidelines to get the output. However, the outcome or final result is not always clear to everyone. Sometimes, the final outcome needs additional explanation or interpretation. The qualitative approach is completely based on non-numerical objects or ranks such as best, better, good, strongly agree, moderately agree, and not agree. This type of assessment approach is beneficial for communicating risk outcomes. One significant drawback of this approach is that the variety of the properties of the factors is very limited; as a result, the comparison and prioritization of the properties and outcomes become difficult. The final approach is the Semi-Quantitative assessment approach, which is somewhere beneficial to both qualitative and quantitative approaches. This approach utilizes the proxy number, ranges, and bin concepts where the values of those concepts are not used in other concepts. The capabilities of translating the scale or bin into qualitative terms and enabling relative comparisons among values at the same time is one of the strongest sides of this approach [10].

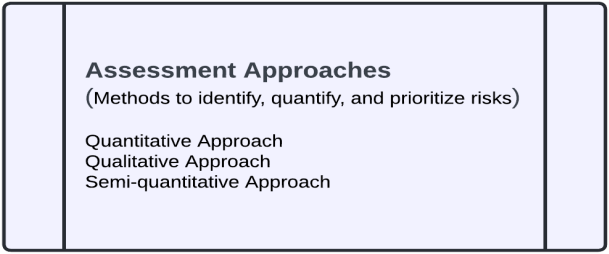

Fig 6: Assessment Approaches [10]

### 5.3.3 Analysis Approaches

The analysis part is essentially the core of risk assessment. The outcome of the risk assessment substantially depends on the quality of the risk analysis. NIST suggests three approaches for risk analysis: i) Threat-oriented approach, ii) Asset-oriented approach, and iii) Vulnerability-oriented approach. All of those approaches are different from each other due to variability factors such as the embarking aspect of risk assessment, degree of specification, and way of treating the threats.

The first priority of the Threat Oriented Approach is to identify the sources of threats and events of threats. This approach also keeps its focus on developing threat scenarios along with identifying the vulnerability of the threat context. Additionally, this approach identifies the effect of adversarial threats based on antagonist intent. The asset-oriented approach or impact-oriented approach identifies the sensitive assets and aftereffects of the concerned area, whereas the vulnerability-oriented approach begins with identifying the set of lacking or limitations of the overall processes. It also determines the events of threats in detail. Each of the approaches considers the same risk factors and, therefore, encompasses similar types of assessment activities yet in distinct orders [10]. Organizations can also use additional rigor analysis approaches such as graph-based analysis, which is different from the three approaches suggested by NIST.

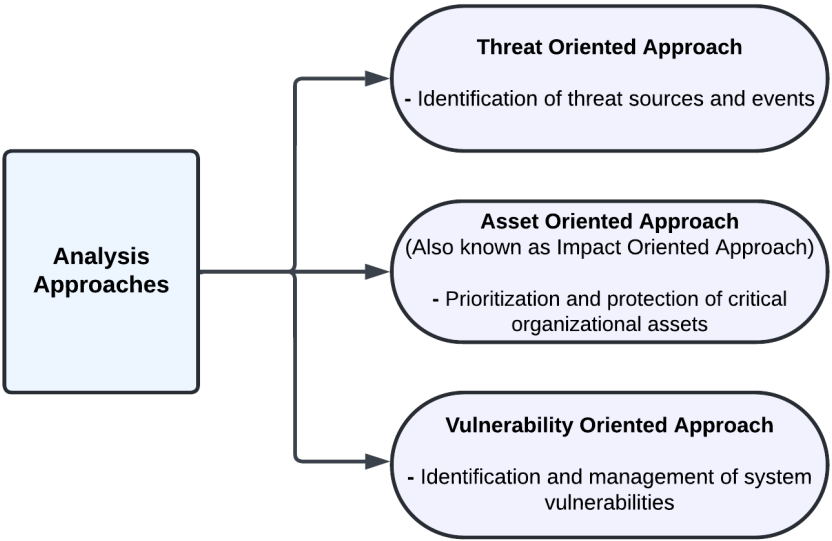

Fig 7: Analysis Approaches for Fundamental Risk Principles [10]

## VI. RISK ASSESSMENT WITH PRACTICAL EVALUATION METRICS

To Support risk assessment and response decisions, the practical evaluation metrics aim to establish a prioritized list of information security threats. In order to accomplish this assessment, business organizations need to conduct risk assessments that consider threats, vulnerabilities, impacts, probabilities, and uncertainties. At this stage, it is necessary to gather all of the necessary data for each assignment. According to the precise definitions, guidelines, and instructions laid out during preparation, the risk assessments should encompass the full scope of the threat landscape. To provide full coverage, it may be necessary to generalize threat sources, threat events, and vulnerabilities in order to provide enough coverage owing to resource restrictions. However, when necessary, extensive assessments of individual sources, events, and vulnerabilities will be done.

This chapter follows the guidelines set by NIST SP 800-30 and ISO 31000:2018 for risk assessment and management. While these standards provide the foundation, each section includes practical enhancements incorporated to meet organizational challenges more effectively. For instance, while NIST suggest methods for assessment scale or level of risk, this chapter introduces a customized scoring system (Risk Criticality Level). This ensures that the evaluations are both thorough and relevant to the organization's specific context. Throughout the chapter, the risk assessment process has been adjusted to be more flexible and actionable, all while maintaining alignment with the principles of NIST and ISO.

The following steps comprise the risk assessment procedure:

**6.1. Identify the Asset/Process/Service to be assessed**

To determine which assets or services should be evaluated in the context of risk management, begin by creating a thorough inventory of all tangible and intangible assets, as well as the associated processes and services. The things in this list should be categorized based on their types, such as hardware, software, or data. After categorization, rank each item based on its importance to business operations and its overall criticality. This prioritization facilitates assessing their susceptibility and the possible consequences of security threats. Ultimately, choose the most vital assets, processes, or services to undergo a thorough evaluation, with a specific emphasis on those that, if undermined, would present the highest level of threats to the business. This methodical technique guarantees a focused and effective procedure for evaluating risks [10] [11].

**6.2. Determine the Asset Value**

In order to ascertain the worth of assets inside an organization, it is vital to engage important entities such as the Organizational IT Security Committee, IT Steering Committee, and Risk Management Committee [28]. Begin by identifying crucial stakeholders from these committees to engage in the valuation process. Thereafter, the criteria for evaluating the value of anything should be established, which should encompass factors such as the financial consequences, legal ramifications, operational effects, and detriment to reputation [10] [11]. Organize review workshops or meetings with these committees to analyze and evaluate the value of each asset according to the defined criteria [10] [11]. Utilize a scoring mechanism, such as a 1-5 scale, to allocate value ratings to each asset, which will reflect the combined input from the committees [10] [11]. Ultimately, it is crucial to record and document these values and carry out regular assessments to guarantee their precision and applicability to the current requirements of the company [10] [11]. This cooperative method guarantees a thorough and balanced assessment of asset worth, which is essential for efficient risk mitigation. Below are some common IT assets, processes, and services to consider, though this list is not limited to these examples.

| Asset/Service | Asset Value for an Asset/Process/Service (1-5) |
|---|---|
| Core Software | This value should be determined by the system owner and the CIO, and then proposed and approved by one of the organization's risk management or security committees. The value must take into account both the tangible and intangible aspects of the asset, service, or process. |
| Primary Database | |
| Network Infrastructure | |
| Cloud Storage Services | |
| Data Center | |
| Business Continuity Solutions (DR/Far DR) | |
| Integrated Security Operations | |
| Payment Processing Systems | |
| Endpoint Device | |
| Web Server | |
| File Server | |
| Email Server | |
| Customer Relationship Management (CRM) Systems | |

Table 1: Asset Value Evaluation Guide [10] [11]

### 6.3. Identify associated threats and their severity levels for each asset

To identify and assess the level of risk for each asset, start by compiling a list of potential threats through methods such as knowledge sessions, analysis of past data, threat modeling, observations from internal and external audits, incident logs analysis, vulnerability assessments, penetration testing, and consultation of industry reports, among other approaches. Enhance this list by incorporating information from threat intelligence platforms and industry-specific databases. Evaluate the seriousness of these threats by considering elements such as monetary loss, disruption of operations, legal consequences, and harm to reputation, utilizing frameworks like NIST SP 800-30 or ISO/IEC 27005 and suitable risk assessment software. Assign a numerical value to each threat based on a predetermined scoring system (such as a scale of 1-5) in order to accurately measure the degree of its impact. Record these discoveries in a consolidated database, utilizing technologies such as risk management software or Excel. Consistently assess and revise the list of potential threats and their levels of severity in order to stay up to date with changes in the threat landscape and the environment in which assets are located. This can be done by scheduled reviews and automated warnings from threat intelligence platforms, ensuring that the risk register remains current. This comprehensive strategy guarantees that risk assessments are exhaustive and that mitigation measures are targeted with great effectiveness [10] [11].

| Level of Threat ||
|---|---|
| **Threat Rating** | **Impact on Business** |
| 1 | Insignificant |
| 2 | Minor |
| 3 | Moderate |
| 4 | Major |
| 5 | Catastrophic |

Table 2: Level of Threat [10]

### 6.4. Identify vulnerabilities and their impact on the CIA

For the evaluation of vulnerabilities and their impacts on the Confidentiality, Integrity, and Availability (CIA) triad is initiated by identifying vulnerabilities using techniques such as vulnerability scanners, penetration testing, and security audits. This should be complemented by examining system logs, configurations, and past incident reports. Subsequently, assess the possible impact of each vulnerability on the organization's confidentiality, integrity, and availability. This involves identifying the possibility of unauthorized disclosure of information, manipulation of data, or disruption of services. Assess the influence of each vulnerability on the CIA triad by utilizing a predetermined scale ranging from 1 (negligible influence) to 5 (highest influence) to assess the seriousness. Record these findings in a centralized repository utilizing risk management software or tools like Excel or Google Sheets for well-organized documentation. Regularly evaluate and revise this evaluation to accommodate alterations in the threat environment, system setups, and security protocols, guaranteeing ongoing safeguarding and preparedness for responding to incidents [10] [11].

| Level of Vulnerabilities ||
|---|---|
| (Impact of VA is specified considering existing vulnerabilities, remediation measures and impact on CIA) ||
| **Vulnerability Rating** | **Impact on Business** |
| 1 | Negligible |
| 2 | Low or Minimal |

| 3 | Medium |
| 4 | High |
| 5 | Highest |

Table 3: Level of Vulnerabilities [10]

### 6.5. Determine Remediation Measures

Determination of remediation measures for vulnerabilities affecting the Confidentiality, Integrity, and Availability (CIA) triad, set in motion by developing targeted remediation strategies, such as software patches, configuration updates, and policy enhancements. Prioritize these actions based on the severity and risk to the organization's critical functions, assessing the required resources like budget, personnel, and time for feasibility. Implement these measures in order of priority and continuously monitor their effectiveness in mitigating the vulnerabilities. Maintain detailed documentation of all remedial actions and outcomes and regularly report to stakeholders to adapt to evolving threats and organizational changes. This structured approach ensures ongoing alignment with security and risk management objectives.

### 6.6. Assess the level and rating of Vulnerability Assessment (VA)

To accurately assess the effectiveness of a vulnerability assessment (VA) post-remediation, begin by re-evaluating each asset's susceptibility to threats, considering updated factors like location, accessibility, and enhanced security measures. Reassess the impact of vulnerabilities on confidentiality, integrity, and availability using a scale from 0 (no impact) to 4 (critical impact), alongside the asset's exposure to threats, now graded from 1 (negligible risk) to 5 (highest risk) to reflect remedial improvements. Combine these updated metrics to form a comprehensive vulnerability rating that captures the residual risks post-remediation. Document and organize these findings in risk management software or spreadsheets, ensuring regular reviews to maintain accuracy against the evolving threat landscape. This structured re-evaluation is crucial for measuring remediation effectiveness and prioritizing future security efforts based on the severity of remaining risks and asset exposure [10] [11]. The study followed the concepts of NIST & ISO and came up with a practical version of overall vulnerability rating post-remediation.

| Vulnerability ID | Asset Affected | Exposure Level | Impact on Confidentiality (0-4) | Impact on Integrity (0-4) | Impact on Availability (0-4) | Overall Vulnerability Rating (1-5) |
|---|---|---|---|---|---|---|
| VULN1 | Web Server | High | 4 | 3 | 0 | The overall score for Confidentiality, Integrity, and Availability (CIA) should be set by integrating individual CIA ratings, reflecting the effectiveness of remediation measures, current security needs and most importantly the Risk Tolerance Level set by the organization. |
| VULN2 | Database Server | Medium | 2 | 4 | 4 | |
| VULN3 | Employee Laptop | Low | 3 | 2 | 3 | |

Table 4: Overall Vulnerability Rating Post-remediation [10] [11]

**6.7. Calculate the threat value, which is the sum of the threat level and vulnerability level**

To estimate the risk, begin by ascertaining the threat level on a scale ranging from 1 to 5, where 1 signifies a low threat and 5 signifies a significant threat. Utilize these evaluations to compute the degree of danger by employing the following equation: Threat Value = Threat Level + Vulnerability Level. After doing the calculations, store these threat values in a centralized repository to guarantee that all data is methodically documented and easily available for continuous risk management.

**6.8. Identify the likelihood of threats**

To assess the probability of attacks, embark on gathering historical data from previous incident reports, threat intelligence feeds, and industry statistics to evaluate the frequency of comparable threats. Assess the existing threat environment by utilizing threat intelligence platforms and security advisories to obtain a full understanding of potential hazards. Assess various criteria, including the level of asset vulnerability, the effectiveness of current security measures, and the capability of possible attackers. Utilize the provided assessments to assign likelihood ratings using a predetermined scale that spans from 1 (indicating rarity) to 5 (indicating a high probability). Record these ratings in a centralized database and implement a regular process for assessing and revising the ratings to ensure they are precise and reflect any alterations in the threat landscape [10] [11].

| Likelihood | |
|---|---|
| **Rating** | **Possibility of Occurring** |
| 1 | Very Unlikely |
| 2 | Unlikely |
| 3 | Possible |
| 4 | Likely |
| 5 | Very Likely |

Table 5: Likelihood of threats in possibility [10]

**6.9. Calculate the Risk Impact Rating by multiplying the Asset Value, Threat Value, and Likelihood**

| Risk Impact Rating |
|:---:|
| Asset Value<br>X<br>Threat Value (Level of Threat + Level of Vulnerability)<br>X<br>Likelihood |

Table 6: Risk Impact Rating calculation [10]

**6.10. Specify the risk criticality level (1- Low, 2- Medium, 3- High, and 4- Critical)**

While these tasks are provided in a sequential manner for clarity, it is important to note that, in reality, some iteration among the tasks is both essential and anticipated. Depending on the specific objective of the risk assessment, the business may discover that rearranging the duties might be advantageous. The risk assessments must conform to the stated objective, scope, assumptions, and limits established by the entity that initiates them, regardless of any modifications made [10] [11]. The criteria outlined below are provided by the 'AssessITS' team; however, they may differ among organizations depending on their individual risk tolerance levels. The Risk Criticality Level framework adapts the risk categorization principles from the NIST 800-30 guide. While NIST utilizes a qualitative and semi-quantitative approach, our model introduces a detailed scoring system (1 to 250) to assign risk levels like Low, Medium, High, and Critical, based on the potential impact and probability of threats. This allows for a more nuanced evaluation of risks, helping organizations prioritize their responses more effectively.

| Risk Criticality Level ||
|:---:|:---:|
| **Rating & Score** | **Possibility of Occurring** |
| 1 (1 to 45) | Low |
| 2 (46 to 99) | Medium |
| 3 (100 to 199) | High |
| 4 (200 to 250) | Critical |

Table 7: Risk Criticality Level [10]

The following Risk Assessment Matrix can be utilized as a guideline for evaluating risks associated with IT assets/services:

| Sl | Asset/Service | Asset/Service Owner | Asset Value (AV) | Threats | Level of Threat | Vulnerabilities | Remediation Measures | Impact On C/ I/ A | | | Level of Vulnerabilities | Threat Value (TV) | Likelihood of Threat (LH) | Risk Impact Rating | Risk Criticality Level |
|---|---|---|---|---|---|---|---|---|---|---|---|---|---|---|---|
| | | | | | | | | C (0-4) | I (0-4) | A (0-4) | | | | | |
| | Any Asset/Service of the Organization | Wing/Team/IT/any department other than IT/ Organnization | Asset value for an asset/service (1-5) | Threats can be identified through knowledge sessions, historical data analysis, VA, PT, SIEM log, system logs, industry reports, etc. | The level of threat: 5 - Catastrophic 4 - Major 3 - Moderate 2 - Minor 1 - Insignificant | Vulnerabiliies can be determined by VA, PT, or by Technical Assessments | Measures taken in response to the vulnerabilities found | C -Confidentiality I - Integrity A - Availability | | | Exposure of Asset: 5 - Highest 4 - High 3 - Medium 2 - Low or Minimal 1 - Negligible | Level of Threat + Level of Vulnerability | Level of Probability/Likelihood: 5 - Very Likely 4 - Likely 3 - Possible 2 - Unlikely 1 - Very Unlikely | (AV x TV x LH) 200 to 250- Critical 100 to 199- High 46 to 99 - Medium 1 to 45- Low | Critical- 4 High- 3 Medium- 2 Low- 1 |

Figure 8: Risk Assessment Matrix [10]

Here is a detailed step-by-step anticipated scenario, thoughtfully designed to align seamlessly with the Risk Assessment Matrix provided above, while drawing on established risk assessment standards [10] [11].

### 6.10.1. Asset Identification and Owner
- Asset/Service: Web Server
- Owner: IT Department

### 6.10.2. Asset Value (AV)
- 4 - Critical for business operations, hosting important applications.

### 6.10.3. Threat Identification
- Includes general cybersecurity threats that could impact the web server, such as unauthorized access, data theft, and service disruption.

### 6.10.4. Level of Threat (TV)
- 4 (Major) - Considering the worst-case scenario for a critical asset like a web server.

### 6.10.5. Vulnerability Assessment
- Identified vulnerabilities include insufficient firewall protection, outdated server software.

### 6.10.6. Remediation Measures
- No remediation measures have been implemented because the company is not prepared to purchase a new firewall. Additionally, updating the server software is not feasible as the current web service running on this server is incompatible with the newer version.

### 6.10.7. Impact on CIA
- Impact on CIA: C - 4, I - 4, A - 4

### 6.10.8. Level of Vulnerabilities
- 5 (Highest), considering identified vulnerabilities, absence of remediation measures, and their impact on CIA

### 6.10.9. Calculation of Threat Value (TV)
- TV = Level of Threat + Level of Vulnerability = 4 + 5 = 9

### 6.10.10. Likelihood of Threat (LH)
- The likelihood is assessed as 4 (Likely) because both the threat and vulnerability levels are high, indicating that the web server is at significant risk of a cyber-attack.

### 6.10.11. Risk Impact Rating (RI) Calculation
- RI = AV × TV × LH = 4 × 9 × 4 = 144

### 6.10.12. Risk Criticality Level
- High (144), The Risk Criticality Level is classified as high. Values ranging from 100 to 199 are categorized as high according to the 'AssessITS' criteria.

## VII. CONCLUSION

This research has illustrated the efficacy of the 'AssessITS' strategy in bridging the divide between theoretical risk assessment principles and their actual implementation in the fields of IT and cybersecurity. 'AssessITS' improves the efficiency of risk assessments and strengthens enterprises' ability to withstand cyber threats by utilizing recognized standards like NIST, COBIT, and ISO. In the future, 'AssessITS' will be well-positioned to adjust to and minimize emerging risks, making it crucial in the ever-changing field of digital security. The continuous advancement and improvement of this approach will be essential for predicting and addressing changing security difficulties, guaranteeing its significance for risk assessment strategy.

## VIII. FUTURE SCOPE

Organizational Information Technology (IT) and Cybersecurity risk assessment are crucial as all business operations are primarily developing or following several risk assessment standards. The future scope of the presented study can be directly related to future cyber threats, possible cyber-attacks, and vulnerabilities discovered (even undiscovered vulnerabilities). Several revisions and updates on risk assessment frameworks seem to be needed in the future as a systematic process to reduce risks. The guidelines for the integration of such approaches are primarily beneficial for several organizational committees, including the Risk Management Committee and IT Security Committee. While 'AssessITS' provides a comprehensive and standardized approach that can be implemented directly, organizations also have the option to customize the framework to better fit their unique risk profiles if needed. Future research will focus on developing a more highly customizable version of the framework, allowing for even greater adaptability to specific operational environments and risk landscapes.

However, organizations operating in highly specialized industries or dealing with unique risk scenarios may find it challenging to adapt the framework without additional customization. Additionally, smaller organizations with limited expertise in risk management might require some initial guidance or external consultation to fully leverage the method's potential. The success of implementation also depends heavily on the availability of accurate data and the expertise of the practitioners using the method. Going forward, the 'AssessITS' approach will need to adapt to technological advancements, including AI and machine learning. Future iterations will incorporate real-time data analytics and refined metrics for risk assessment, further customizing the framework for sectors like healthcare, finance, and government, each with specific IT security needs.

**Author Contributions:** Conceptualization - MMR, NK, Methodology - MMR, NK, SAS, MMR, Writing and Editing - MMR, NK, SAS, MMR. All authors have read and agreed to publish the manuscript.
**Funding:** The authors receive no external funding for this research work.
**Data Availability Statement / Consent Statement:** Not applicable.
**Conflicts of Interest:** The authors declare no conflicts of interest.